\documentclass{ws-procs9x6}
\usepackage{latexsym}
\usepackage{graphicx}
\begin{document}

\title{Threshold Resonances - a Lattice Perspective}

\author{D.G.~Richards}

\address{Jefferson Laboratory, MS 12H2,\\ 12000 Jefferson Avenue,\\
Newport News, VA 23606, USA\\
E-mal: dgr@jlab.org}

\maketitle

\abstracts{The calculation of the masses of the lightest nucleon
resonances using lattice QCD is surveyed.  Recent results for the mass
of the first radial excitation of the nucleon, the Roper resonance,
are reviewed and the interpretation in terms of models of hadronic
resonances, such as the quark model and hadronic molecules, discussed.
The talk concludes with an outline of prospects for future
calculations.}

\section{Introduction}
The calculation of the light hadron spectrum has historically been the
benchmark calculation of lattice QCD, but the predictive focus has
been on the phase structure of QCD and on weak matrix elements.  There
is now increasing interest given to the study of hadronic structure,
both through the measurement of form factors and structure functions,
and through the determination of the hadron spectrum, and in
particular the nucleon resonance spectrum.  In this talk, I will
review recent lattice results, emphasising what lattice QCD
measurements can tell us about the threshold resonances.

I will begin by briefly outlining the theoretical and computational
issues in the determination of the nucleon spectrum.  I will then
review recent lattice results, before addressing the question of what
they can tell us about the nature of the observed threshold states, in
particular by emphasising the importance of continuing the studies to
physical values of the light-quark masses, and by the pursuing the study of
``molecular'' states.  I will conclude by discussing prospects for
future calculations.

\section{Nucleon Spectrum Cookbook}
The recipe for determining the mass of a low-lying state from
Euclidean lattice QCD is straightforward: choose an operator $O_N$
having a large overlap with the state, and form
the time-sliced correlator $C(t)$:
\begin{eqnarray}
C(t) & = & \sum_{\vec{x}} \langle 0 \mid O_N (\vec{x},t)
\overline{O}_N(0) \mid 0 \rangle \\
& \stackrel{\longrightarrow}{t \rightarrow \infty} & e^{-M t}
\end{eqnarray}
where $M$ is the mass of the lightest particle in that channel.  In
practice, there are systematic uncertainties that have to be accounted
for: finite-size and discretisation effects, the extrapolation to
physical values of the light quark masses, and, until recently, the
systematic uncertainty due to the use of the quenched approximation.
For the light-hadron spectrum, there have been many precise
calculations of the lowest-lying states both in the quenched
approximation, and in ``full'' QCD; in the former case, the measured
spectrum agrees with experiment to around 10\% in most
channels\cite{cppacs}.

The calculation of the nucleon resonance spectrum has particular
delicacies.  Firstly, the cubic symmetry group of the lattice limits
our ability to construct interpolating operators of a specific
$J^{PC}$; only the lightest states for $J= 1/2, 3/2~\mbox{ and}~ 5/2$
can be delineated unambiguously from a single correlator.  Secondly,
the propagation of heavier states is subject to worsening
signal-to-noise ratios.  Finally, the determination of the higher
resonances within a channel requires the determination of a matrix of
correlators.

\begin{figure}[ht]
\begin{center}
\epsfxsize=6.5cm   
\epsfbox{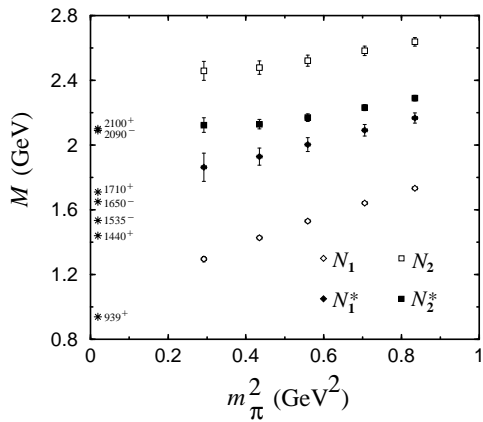}
\end{center}
\caption{Spectrum of the lightest positive- and negative-parity
  nucleon resonances
  using the operators $N_1$ and $N_2$ of
Eq.~(\protect\ref{eq:ops})
\protect\cite{adelaide02}.}
\label{fig:adelaide}
\end{figure}
There have been several studies of the excited nucleon
spectrum\cite{lee98,bnl02,lhpc02,adelaide02}, focusing on the ground states of both
parities for spin-$\frac{1}{2}$ and spin-$\frac{3}{2}$.  All of these
studies have also extracted a radial excitation in the $J^P =
\frac{1}{2}^+$ channel by employing two interpolating operators:
\begin{eqnarray}
N_1 & = & (u C \gamma_5 d) u\nonumber\\
N_2 & = & (u C d) \gamma_5 u.\label{eq:ops}
\end{eqnarray}
These both have an overlap onto $\frac{1}{2}^+$ states, but $N_2$
vanishes in the non-relativistic limit and is expected to couple
predominantly to the radial excitation of the nucleon; a sample
calculation is shown in Figure~\ref{fig:adelaide}.  All the
calculations share the feature that the ordering of the states is
broadly in accord with quark-model expectations, with, for example,
the ``Roper'' in excess of $2~{\rm GeV}$, and no evidence for a light
$\Lambda(1405)^-$.

\section{Light quarks, and more of them.}
The preceding calculations have several limitations.  Firstly, they were
all obtained in the quenched approximation to QCD.  Secondly they
employed quarks with masses around that of the strange quark.
Finally, they used a limited basis of operators, and in particular
operators that would be expected to couple primarily to three-quark
states.

The elimination of the quenched approximation to QCD imposes the
greatest computational demands.  It is thus crucial to extract the
maximum physical information; the use of Bayesian statistics
with suitable priors provides a possible means of so doing.  An
application of this method to the study of the nucleon correlator,
albeit for quark masses around the $s$-quark mass, reveals the same
quark-model-like ordering of states observed in the quenched
approximation\cite{maynard02}.

\begin{figure}[ht]
\begin{center}
\epsfxsize=6cm   
\epsfbox{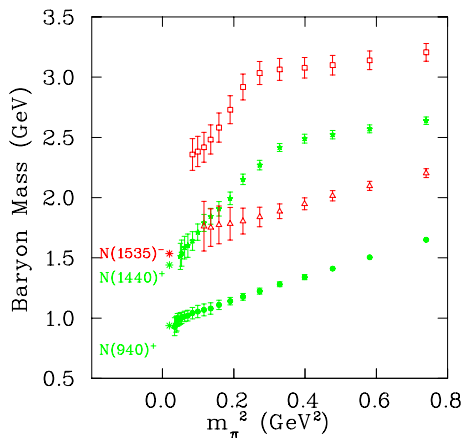}
\end{center}
\caption{Masses of the ground and first-excited states of positive and
  negative parity, obtained from a Bayesian fit to
  $N_1$\protect\cite{fxlee02}.}
\label{fig:uky}
\end{figure}
The discovery of a lattice fermion action possessing an exact chiral
symmetry has opened the prospect for calculations at physical
light-quark masses, though at consideratbly computational cost than
with ``traditional'' Wilson fermions.  A recent calculation using the
overlap realisation of this action has allowed pion masses as low as
$180~{\rm MeV}$ to be attained, enabling the exploration of the region
in which the effects of the pion cloud can emerge\cite{fxlee02}.
Bayesian statistics are used, and a fit to the nucleon correlator
reveals a dramatic cross-over between the lightest negative-parity
state and the radial nucleon excitation for quark masses around
$300-400~{\rm MeV}$, as shown in Figure~\ref{fig:uky}; the
experimental ordering of the states is observed, with the correct
ordering also appearing in the $\Lambda$ channel.

The authors are taking care to investigate the possible sources of
systematic uncertainties in their calculation.  Most notably, even in
the quenched approximation to QCD, both the radial excitation of the
nucleon and its parity partner can ``decay'' for suitably light pion
masses.  In the scalar-meson sector, this is manifest through
non-unitary behaviour of the scalar propagator, which can be
understood within quenched chiral-perturbation theory.\cite{bardeen02}
The final interpretation of the results in the nucleon sector must await
this analysis.

If the experimentally observed spectrum proves to defy interpretation
from these calculations, is it possible that we can interpret the
anomalously light components of the spectrum, such as the Roper and
$\Lambda(1405)^-$, in terms of molecular or multiquark states?  All
calculations so far have used local, three-quark interpolating
operators, and are therefore sensitive to states having broadly that
structure.  Therefore I will conclude this talk with the issue of how
we might observe molecular states.

Typically the binding of hadronic ``molecules'' is small on the scale
of QCD, of the order of a few MeV, and furthermore the states are
large on the scale of the box sizes in current lattice calculations.
The most extensively studied multiquark state has been the H
dibaryon, a proposed six-quark state and the lightest possible spin-0 state
with strangeness -2.  This was originally computed in the bag model to
be $O(100)~{\rm MeV}$ below the $\Lambda \Lambda$ threshold. This
relatively large predicted binding energy has encouraged several
lattice studies over the past 15
years\cite{pbm85,iwasaki88,pochinsky,wetzorke02}, but with conclusions
complicated by the need to extrapolate to infinite volume.  The most
recent study provides no evidence for binding in the infinite-volume
limit\cite{wetzorke02}.

A very fruitful arena in which to explore binding of molecular states
in QCD is the heavy-quark sector, and in particular the $B B$
system\cite{dgr90}.  By using static heavy quarks, the two ``atoms''
are fixed in space, and an adiabatic potential can be defined between
them which can then be probed; here there is evidence of a binding
potential in some channels\cite{michael99,fiebig02}, but models are
required to extend the analysis to physical quark masses.

An elegant way of exploring the issues of scattering lengths
and hadronic interactions is provided by examining the volume
dependence of the two-particle spectrum\cite{luscher86}.  The
method typically requires the computation of
all-to-all propators.  Whilst such calculations are computationally very
expensive for QCD, the advent of terascale computing resources should
enable the tackling of these challenging problems over the next few years.

This work was supported in part by DOE contract DE-AC05-84ER40150
under which the Southeastern Universities Research Association (SURA)
operates the Thomas Jefferson National Accelerator Facility, and by
DE-FG02-97ER41022.  I am grateful to Robert Edwards, George Fleming
and Frank Lee for many helpful conversations.


\begin{thebibliography}{0}
\bibitem{cppacs} S.~Aoki \textit{et al.} (CP-PACS Collaboration),
  hep-lat/0206009. 
\bibitem{lee98} F.X.~Lee and D.B.~Leinweber, Nucl.\ Phys.\ B (Proc.\
  Suppl.) 74, 258 (1999).
\bibitem{bnl02} S.~Sasaki, T.~Blum and S.~Ohta, Phys.\ Rev. D65, 074503 (2002).
\bibitem{lhpc02} M.~G\"{o}ckeler \textit{et al.} (LHPC/QCDSF/UKQCD
Collaborations), Phys.\ Lett.\ B532, 63 (2002).
\bibitem{adelaide02} W.~Melnitchouk \textit{et al.}, hep-lat/0202022.
\bibitem{maynard02} C.M.~Maynard and D.G.~Richards, hep-lat/0209165,
  Nucl.\ Phys.\ B (Proc.\ Suppl.), in press.
\bibitem{fxlee02} F.X.~Lee \textit{et al.}, hep-lat/0208070, Nucl.\
  Phys.\ B (Proc.\ Suppl.), in press.
\bibitem{bardeen02} W.~Bardeen \textit{et al.}, Phys.\ Rev.\ D65,
  014509 (2002).
\bibitem{pbm85} P.B.~Mackenzie and H.B.~Thacker, Phys.\ Rev.\ Lett.\
55, 2539 (1985).
\bibitem{iwasaki88} Y.~Iwasaki \textit{et al.}, Phys.\ Rev.\ Lett.\
  60, 1371 (1988).
\bibitem{pochinsky} A.~Pochinsky \textit{et al.}, Nucl.\ Phys.\
  (Proc.\ Suppl.) 73, 255 (1999).
\bibitem{wetzorke02} I.~Wetzorke and F.~Karsch, hep-lat/0208029, Nucl.\
  Phys.\ B (Proc.\ Suppl.), in press.
\bibitem{dgr90} D.G.~Richards, D.K.~Sinclair and D.~Sivers, Phys.\ Rev.\
  D42, 3191 (1990).
\bibitem{michael99} C.~Michael and P.~Pennanen, Phys.\ Rev.\ D60,
  054012 (1999).
\bibitem{fiebig02} M.S.~Cook and H.R.~Fiebig, hep-lat/0210054.
\bibitem{luscher86} M.~L\"{u}scher, Comm.\ Math.\ Phys.\ 104, 177
  (1986); 105, 153 (1986).
\end{thebibliography}
\end{document}